\documentclass[reprint,prl,showpacs,letter]{revtex4-1}
\usepackage{bm}
\usepackage{graphicx}
\usepackage[usenames]{color}
\usepackage{amsmath,amsfonts,amssymb}
\usepackage{hyperref}
\usepackage{txfonts}

\begin{document}

\title{Fractionally Quantized Berry Phase, Adiabatic Continuation, %Limit
 and Edge States}

\date{\today}
\author{Toshikaze Kariyado}\email{kariyado@rhodia.ph.tsukuba.ac.jp}
\author{Yasuhiro Hatsugai}\email{hatsugai@rhodia.ph.tsukuba.ac.jp}
\affiliation{Division of Physics, Faculty of Pure and Applied Sciences,
University of Tsukuba, Tsukuba, Ibaraki 305-8571, Japan}
% 73.20.-r: Surface states
% 03.65.Vf: Berry phase & topological phases
% 03.65.Ud: Quantum entanglement

\begin{abstract}
 Symmetry protected quantization of the Berry phase is discussed in
 relation to edge states. Assuming an existence of some adiabatic
 process which protects quantization of the Berry phase, non trivial
 Berry phase $\gamma=\pm 2\pi\rho$ ($\rho$ is a local filling of
 particles) for the bulk suggests appearance of edge states with
 boundaries. We have applied this generic consideration for Bloch states
 of some two dimensional model with massless Dirac fermions where
 $\gamma=\pm\pi/2$ implies the edge states. Entanglement entropy is
 evaluated for the models and its relation to the bulk--edge
 correspondence of Dirac fermions is discussed as well.
\end{abstract}

\maketitle

Characterization of phases is one of the main targets of
condensed matter physics. As for a description of physical states, 
roles of boundary conditions have been assumed secondary. 
It is true when one considers classical order using a 
order parameter in the  thermodynamic limit,
since the dimensions of boundaries are less than the
bulk dimension. In topological phases characterized by absence of local
order parameters\cite{PhysRevB.40.7387}, the situation is different. 
In contrast to the symmetry broken phases
with low energy excitations as the Nambu-Goldstone boson,
 the ground state of the topological phase is mostly gapped as a bulk. 
With boundaries or impurities, however, 
there exist low energy excitations only near such geometrical
disturbance\cite{PhysRevB.25.2185,PhysRevLett.42.1698,PhysRevLett.72.1526,PhysRevLett.74.3451,JPSJ.65.1920,1063-7869-44-10S-S29,volovik}.
This edge states/boundary states characterize the topological phases.
For instance, the number of edge
states of quantum Hall system can be predicted by Chern number defined
with the bulk Hamiltonian\cite{PhysRevLett.49.405}, and this is known as
the bulk--edge correspondence\cite{PhysRevLett.71.3697,PhysRevB.48.11851}.
Another example is $Z_2$ topological
insulator\cite{RevModPhys.82.3045,RevModPhys.83.1057} that is
characterized by nontrivial $Z_2$ topological number and
exhibits novel surface states.

In order to analyze the origin of edge states, an adiabatic continuation
is useful. Assuming a modification of the gapped ground state of the bulk to 
a simple state
without gap closing, one may reduce the topological properties of the 
physical system to those of the simple one.
When the reduced system is composed of
independent clusters, edge states can be understood as dangling states
appearing as a result of breaking a cluster at the generic edges\cite{1063-7869-44-10S-S29,PhysRevLett.89.077002}. On the other
hand, if the boundary does not break a cluster,
i.e., the boundary is in between two adjoined clusters, there is no
obvious reason to have edge states. The adiabatic continuation 
is  more powerful when it is combined with 
topological quantities defined by the Berry
connection \cite{Berry84,doi:10.1143/JPSJ.73.2604,doi:10.1143/JPSJ.74.1374,doi:10.1143/JPSJ.75.123601} 
using  gauge twists as parameters. 
Among the topological quantities, the Chern numbers are 
quantized by its definition but the Berry phase 
and its generalizations are quantized with the help
of some
symmetries\cite{PhysRevLett.89.077002,Hatsugai10Z2,PhysRevB.88.245126}. When
such symmetries present, the Berry phase based argument is robust
against adiabatic continuation, as far as the symmetry is kept during
the adiabatic continuation. Furthermore, the entanglement entropy 
is also useful to characterize topological
properties and edge states\cite{PhysRevB.73.245115,PhysRevB.79.205107,0295-5075-95-2-27003,PhysRevB.83.245132,PhysRevB.84.205136}. 

In this paper, we first give general arguments for characterizing a
gapped and short-range entangled state using an adiabatic continuation
and the Berry phase. A natural interpretation of the
bulk--edge correspondence in that general framework is also given.
Then to demonstrate the general idea, we introduce a model
having Dirac cones in its bulk energy dispersion as an example where an
unusual type of the quantization of the Berry phase, the quantization into
$\pm\pi/2$, or fractional quantization is used 
to demonstrate the bulk-edge correspondence of the
Dirac fermions. 

\begin{figure}[tb]
 \begin{center}
  \includegraphics[scale=1.0]{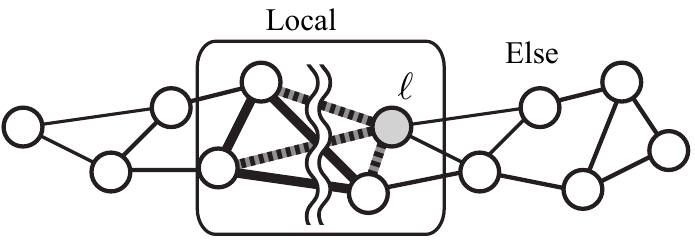}
  \caption{General idea of the Berry phases and  edge states. Hopping of the
thick  dotted lines are modified. When the local object $L$ is decoupled 
(only the hopping on the thick lines is non zero ), this gauge twists are 
gauged away.
}
  \label{fig1}
 \end{center}
\end{figure}

Let us start our discussion from a generic lattice model of spinless
fermions by a hamiltonian with an adiabatic parameter $\lambda $ 
(extension with spins or for systems with $U(1)$ gauge invariance is
straightforward)
\begin{align*} 
H(\lambda ) &= H_{E,E}+H_{L,L}+ \lambda H_{L,E}
\\
H_{\alpha \beta } &= \sum_{i\in \alpha,j\in \beta   }(
c_i ^\dagger t_{ij}
c_j+h.c.+ V_{ij} n_i n_j),\ \alpha, \beta  ={L,\ E}
\end{align*}  
where 
 $n_i=c_i ^\dagger c_i$ and 
the system is divided into two parts $L$ and $E$. 
The parameter $\lambda $ 
is a coupling between them 
(Fig. \ref{fig1}).
It is invariant for a $U(1)$ gauge transformation $H\to H ^\prime =H$ where
$c_i \to c_i ^\prime =\Omega _i c_i$
and $t_{ij}\to t_{ij}^\prime =
\Omega _i  t_{ij}\Omega ^{-1}  _j$,  ($|\Omega _i|=1$). 
We assume the manybody ground state $|G(\lambda ) \rangle $ of $H(\lambda )$ is
always
 gapped for $0 \le \lambda \le 1$.  
The physical ground state $|G(1) \rangle $ is
adiabatically connected to the decoupled ground state $|G(0) \rangle $
 written as
\begin{equation}
|G(0) \rangle =
\sum_{i_1 i_2\cdots \in\, L}\psi^{L}_{i_1i_2\cdots}c_{i_1} ^\dagger c_{i_2} ^\dagger \cdots
\sum_{j_1 j_2\cdots \in\, E}\psi^{E}_{j_1j_2\cdots}c_{j_1} ^\dagger c_{j_2} ^\dagger \cdots
 | 0 \rangle.
\end{equation} 
It implies that the ground state $|G(1) \rangle $ is short-range entangled,
that is, the ground state is composed of local quantum objects. 
Typical such examples are the Haldane phase of the spin 1 chain and
the valence bond solid
states\cite{doi:10.1143/JPSJ.75.123601,PhysRevB.82.155138,PhysRevB.85.075125}.
To characterize this short-range entanglement, let us define
a Berry phase  $\gamma$
by introducing a gauge twist $\omega_\ell =e^{\mathrm{i}\theta}$, 
($0\le\theta\le 2\pi$) at some sites $\ell$'s inside 
the local object $L$ ($\ell$'s  can be multiple sites). 
This twist dependence is given 
by the local hamiltonian $H_{L,L}(\omega_\ell )$
where the hopping $t_{i\ell}$ connecting  the 
sites $\ell$'s and the
remained sites $i$'s inside the local object $L$ are replaced by
$t_{i\ell}\omega_\ell $. 
Note that this gauge twist does not affect any coupling in $H_{L,E}$. 
Then the  Berry phase  defined below characterizes the locality of the
gapped phase
\begin{equation}
\mathrm{i}\gamma = \int \langle G| d G \rangle=\int_0^{2\pi} \mathrm{d}\theta\, 
 \langle G| \partial _\theta G \rangle\quad {\rm mod}\, 2\pi.\label{Berry_def}
\end{equation} 
As for the decoupled case, 
this $\gamma $ is easily evaluated as $\gamma = 2\pi \bar\rho$ 
where 
$\bar \rho=\sum_\ell(2\pi) ^{-1} \int_0^{2\pi} d\theta\, \langle G(0)|n_\ell | G(0)\rangle $ 
 is sum of an averaged filling of the fermion 
at the site $\ell$ since the gauge twist 
$\omega _\ell$ is gauged out by the gauge 
transformation\cite{PhysRevB.78.054431}.
If there exists some symmetry (such as the chiral symmetry, reflection,
time reversal) to guarantee the quantization of the Berry phase $\gamma$, 
this Berry phase is an adiabatic invariant and used for a topological
order parameter at the physical  point $\lambda =1$. 
We have many successful examples for such
situations\cite{doi:10.1143/JPSJ.75.123601,doi:10.1143/JPSJ.76.113601,PhysRevB.77.094431,PhysRevB.79.115107}.
Even if such a symmetry is absent, we may still expect that substantially 
large value of  $\gamma $ implies an existence of 
the short range entanglement. That is, the Berry phase $\gamma $ can be
a good topological order parameter. 

For the gapped phase that can be well described
by a collection of local objects $L$, a finite $\gamma$ suggests
appearance of edge states when the boundary is on the gauge twisted
bonds. In the decoupled limit, such boundary breaks a local
object and broken pieces appear as edge states\cite{footnote1}. Even for
a finite coupling, the edge states can be still localized with the
symmetry protection, since the Berry phase $\gamma$ is an adiabatic
invariant and the locality of the ground state retains as well.
This general idea can be applied for several systems.
Application is not limited to one-dimensional systems
and  applicable to higher dimensions.
When the system is free of manybody interactions, 
momenta parallel to a given edge (or surface) are regarded  as 
parameters determining effective one-dimensional model.
One of such important examples is
 a zero mode edge state  at the zigzag boundary of
graphene which is characterized 
by the Berry phase in the effective one-dimensional model
\cite{PhysRevLett.89.077002,Hatsugai20091061,PhysRevB.84.195452}.
Quantization of the Berry phase
$\gamma/\pi \in\mathbb{Z}$ is well known today but here in this paper, 
we demonstrate  $\gamma=\pm\pi/2$, i.e, a
fractional quantization of $\gamma$ is
also useful for the bulk-edge correspondence of
the Dirac fermions.

The locality of the gapped ground state is also reflected in
the entanglement entropy. In the $\lambda=0$ limit, there is
no need to consider the entanglement between $L$ and $E$. Then, we
divide $L$ into two parts $L_A$ and $L_B$ (corresponds to breaking the
local object $L$), and calculate the entanglement entropy by
tracing out the information in $L_B$. For the noninteracting case with
one fermion in $L_A$ or one lattice site in $L_A$, the entanglement
entropy is readily evaluated as 
\begin{equation}
 S=-[\bar{\rho}\log\bar{\rho}+(1-\bar{\rho})\log(1-\bar{\rho})],
  \label{EE_formula}
\end{equation}
where $\bar{\rho}$ is the fermion filling in $L_A$. Just as in the case
of the Berry phase, it is determined from $\bar{\rho}$ only, i.e., there
is a solid relation between the Berry phase and the entanglement entropy
in some limit\cite{PhysRevB.73.245115}. $S$ in Eq.~\eqref{EE_formula}
takes maximum when $\bar{\rho}=0.5$, which
corresponds to $\gamma=\pi$.
Although Eq.~\eqref{EE_formula} is for a specific
limit, one may expect strong correlation between the Berry phase and the
entanglement entropy 
through edge states in
general\cite{PhysRevB.77.094431}.

\begin{figure}[tb]
 \begin{center}
  \includegraphics[scale=1.0]{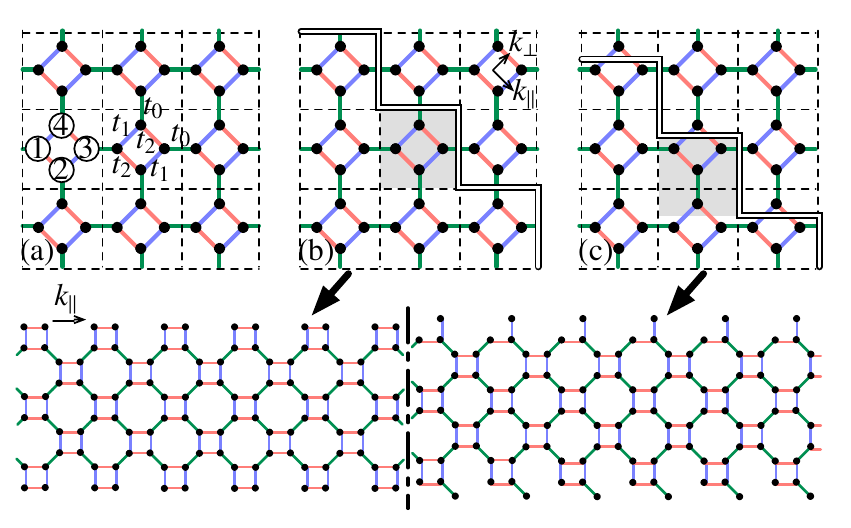}
  \caption{(a) Definitions of transfer integrals. Namings of sublattices
  are also shown. (b) and (c)
  Definitions of type 1 and type 2 edges. Shaded regions are
  corresponding unit cells. Examples of ribbons for edge spectrum
  calculations are shown in lower panels.}
  \label{fig2}
 \end{center}
\end{figure}
As an example to describe general ideas explained above, we introduce a
spinless fermionic tight-binding model
having four sublattices in a unit cell. [See Fig.~\ref{fig2}(a).] We
define three kinds of transfer integrals, $t_0$, $t_1$, and $t_2$ as in
Fig.~\ref{fig2}(a), and set $t_0=1.0$, $t_1=0.5$, and
$t_2=1.5$ throughout this paper. Properties of the very similar
model have been addressed in Ref.~\onlinecite{PhysRevB.88.195104}
very recently, and thus, we concentrate on the Berry phase and its
fractional quantization in this paper. 
For the edge state characterization, we consider two kinds of edge
shapes named as type 1 and type 2. [See Figs.~\ref{fig2}(b).] In order
to treat a two-dimensional model, we
introduce momenta parallel ($k_\parallel$) and perpendicular ($k_\perp$)
to the edge whose directions are shown in Fig.~\ref{fig2}(b), with 
$k_\parallel$ acting as a parameter determining an effective
one-dimensional model. In the present case, the gauge twist
$\omega_\ell=\mathrm{e}^{\mathrm{i}\theta}$ can be regarded as a
twisted boundary condition, and the integration over $\theta$ can be
mapped to the integration over $k_\perp$. Then, the Berry phase
Eq.~\eqref{Berry_def} is given by the Zak
phase\cite{PhysRevLett.62.2747}
\begin{equation}
 \mathrm{i}\gamma(k_\parallel)=\sum_{n\in\text{filled}}
  \int^\pi_{-\pi}\mathrm{d}k_\perp
  \langle u_{nk_\perp k_\parallel}
  |\partial_{k_\perp}|u_{nk_\perp k_\parallel}\rangle,
  \label{Berry_def_mom}
\end{equation}
and we use this expression for computational convenience. Here,
$k_\perp$ is properly scaled so that the Bloch wave function
$|u_{nk_\perp k_\parallel}\rangle$ has periodicity
$2\pi$ in $k_\perp$. 
Note that a unit cell convention is directly related to $\gamma(k_\parallel)$.
For each boundary shape, we set a unit cell so that the given
boundary lies in between two neighboring unit
cells\cite{PhysRevB.88.245126} to use the Berry phase
$\gamma(k_\parallel)$ to discuss the edge states. [See
shaded regions in Figs.~\ref{fig2}(b) and (c).]
We employ the technique in
Refs.~\onlinecite{PhysRevB.47.1651} and \onlinecite{PhysRevB.78.054431}
for the Berry phase calculations. Edge spectrum is calculated with
ribbon geometry like those in the lower panels of Fig.~\ref{fig2},
though the actual calculations are performed on much wider ribbons.
In order to discuss the fractional quantization, we focus on the quarter
filling case, the case of one
fermion per a unit cell, throughout this paper. 

\begin{figure}[tb]
 \begin{center}
  \includegraphics[scale=1.0]{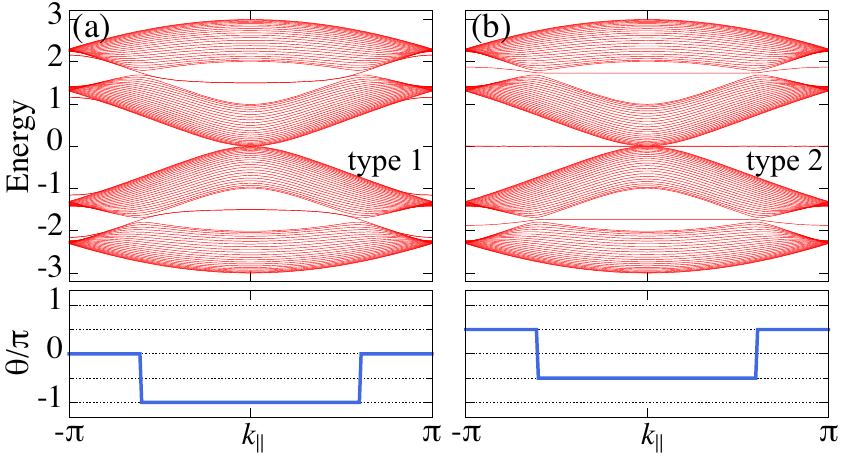}
  \caption{Energy spectra and Berry phases for quarter filling for type
  1 (a) and type 2 (b) edges.} 
  \label{fig3}
 \end{center}
\end{figure}

Figure~\ref{fig3} shows the edge spectra (upper panels) and the Berry
phases (lower panels) for the type 1 (a) and type 2 (b) edges. As we
handle the quarter filling case, we should focus on
the lowest band and the gap just above it. For the given parameter set,
Dirac cones appear between the lowest
and second lowest bands in the bulk energy dispersion. Consequently, 
the bulk continuum, the region filled with bands with bulk nature,
touches at two points, which
are projected Dirac cones. For the type 1 edge, the Berry phase is
quantized into 0 and $\pi$. On the other hand,
for the type 2 edge, the Berry phase is quantized into $\pm\pi/2$,
i.e., the fractional quantization is really achieved. As we have noted,
the Berry phase is related to the site resolved filling
$\bar{\rho}$. Here, the sublattice and $k_\parallel$
resolved filling $\rho_a(k_\parallel)$ plays a role of
$\bar{\rho}$. For the present model, the time reversal symmetry
gives $\rho_a(-k_\parallel)=\rho_a(k_\parallel)$, the inversion symmetry
gives $\rho_1(-k_\parallel)=\rho_3(k_\parallel)$ and
$\rho_2(-k_\parallel)=\rho_4(k_\parallel)$, and the mirror symmetry gives
$\rho_1(-k_\parallel)=\rho_2(k_\parallel)$ and
$\rho_3(-k_\parallel)=\rho_4(k_\parallel)$. 
In addition, $\sum_{a}\rho_a(k_\parallel)=1$ for the quarter
filling. Combining these relations, we finally
obtain the relation $\rho_a(k_\parallel)=1/4$, which leads to the
fractional quantization of the Berry phase into $\pm\pi/2$. 
As this derivation shows, the $\pm\pi/2$ quantization is
a kind of symmetry protection by crystal symmetries.

For the type 1 edge, an edge state distinct from the bulk continuum is
existing as they connect two projected Dirac
cones. In the region with the edge states, $\gamma(k_\parallel)$
takes a value of $\pi$ ($-\pi$ is equivalent to $\pi$). 
Existence/absence of the edge state are
switched at the gap closing point, which is consistent with 
$\pi$-jump in $\gamma(k_\parallel)$\cite{PhysRevB.88.245126}. In this
case, the edge states are
doubly degenerate and they are localized at left and right boundaries,
respectively. In contrast, edge states appearing for the type 2 edge,
for which $\pm\pi/2$ quantization takes place, are different. There
appears only one nondegenerate edge state through the entire edge
Brillouin zone. In this case, the spacial position of the edge states
are switched at the 
projected Dirac cone. Namely, the edge state near
$k_\parallel=0$ lives on the edge at the one side of the ribbon, say
the left edge, while the one near $k_\parallel=\pi$ lives on the edge at the
other side.

\begin{figure}[tb]
 \begin{center}
  \includegraphics[scale=1.0]{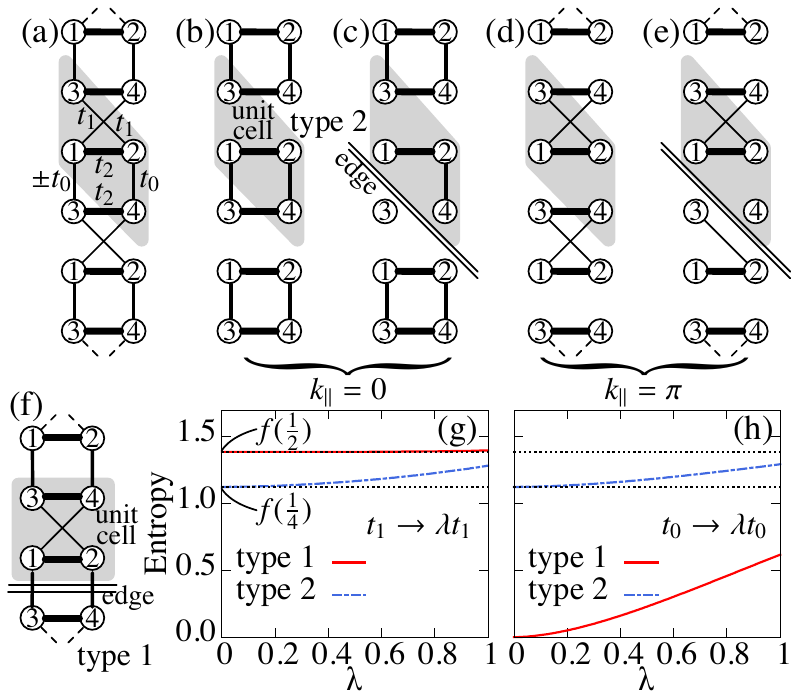}
  \caption{(a) One dimensional model for a fixed $k_\parallel$. For the
  transfer integral indicated as $\pm t_0$, take $+t_0$ ($-t_0$) for
  $k_\parallel=0$ ($\pi$). (b) and (d) Adiabatically continuated
  models. (c) and (e) Adiabatically continuated model after edge
  introduction. (b) and (c) are for $k_\parallel=0$, while (d) and (e)
  are for $k_\parallel=\pi$, respectively. (g) and (h) Entanglement
  entropy for $k_\parallel=0$ (g) and $k_\parallel=\pi$ (h). A function
  $f(x)$ is defined as $f(x)=-2[x\log x+(1-x)\log(1-x)]$.}
  \label{fig4}
 \end{center}
\end{figure}
Here, we apply adiabatic continuation focusing on the
effective one-dimensional model for $k_\parallel=0$ and
$\pi$. Schematically, this one-dimensional
model is illustrated in Fig.~\ref{fig4}(a), in which a
plus (minus) sign should be taken for the transfer integral denoted as
$\pm t_0$ for
$k_\parallel=0$ ($k_\parallel=\pi$). The type 1 and 2 edges and
the corresponding unit cell conventions in the effective one-dimensional
model are indicated as doubled lines and shaded regions in Fig.~\ref{fig4}.
For $k_\parallel=0$, the model can be adiabatically connected, i.e.,
smoothly transformed without closing the gap between the lowest and
second lowest bands, to the model in Fig.~\ref{fig4}(b) by replacing
$t_1$ by $\lambda t_1$ and reducing $\lambda$ from 1 to 0
gradually. Note that the model in $\lambda\rightarrow 0$ limit is
composed of decoupled clusters, or
local objects, and this operation maintains
the symmetry of the model and the Berry phase quantization. 
For $k_\parallel=\pi$, a different adiabatic continuation must be
applied to keep the gap, namely, we should replace $t_0$ by
$\lambda t_0$ and take $\lambda\rightarrow 0$ limit. This operation
results in a model in Fig.~\ref{fig4}(d).

Next, we explicitly show that edge states are induced by breaking local
objects. For $k_\parallel=0$, the energy levels in the
$\lambda\rightarrow 0$ 
limit are $t_2+t_0$, $t_2-t_0$, $-t_2+t_0$, and
$-t_2-t_0$. Then, if the type 2 edge is introduced here, it breaks the
local object at the edge as Fig.~\ref{fig4}(c), and modifies the
energy levels to $\pm(t_2^2+t_0^2)^{1/2}$ and 0 (doubly degenerate).
Since $-t_2-t_0<-(t_2^2+t_0^2)^{1/2}<-t_2+t_0$, the state with
energy $-(t_2^2+t_0^2)^{1/2}$ appear as an ingap edge state in
Fig.~\ref{fig4}(b) near
$k_\parallel=0$. The wave function
for this ingap state has its weight only in the {\it upper} side of the
doubled line in Fig.~\ref{fig4}(c). In the
exactly same way, the origin of the edge state near $k_\parallel=\pi$ in
Fig.~\ref{fig3}(b) can be identified, but due to the difference in the
adapted adiabatic continuation, the wave function for the ingap state
has its weight only in the {\it lower} side of the doubled line in
Fig.~\ref{fig4}(e), which is opposite from the case of $k_\parallel=0$.

Figure~\ref{fig4} also shows the entanglement entropy for
$k_\parallel=0$ and $\pi$. In practice, the
entanglement entropy is numerically
calculated using the formula based on the correlation function 
$\langle c^\dagger_i c_j\rangle$\cite{1751-8121-42-50-504003}, and arranging the effective
one-dimensional model in a
closed circle shape and inserting two cuts to perform bipartition. Here,
two cuts are required to divide a closed circle into
two parts, and two local objects (in the decoupled limit) are broken in
this procedure. We consider two kinds of cutting shapes corresponding to
the type 1 and 2 boundary. In order to see effects of the adiabatic
continuation, we plot $\lambda$ dependence of the entanglement
entropy for $k_\parallel=0$ [Fig.~\ref{fig4}(g)] and $\pi$
[Fig.~\ref{fig4}(h)].
For $k_\parallel=0$, the entanglement entropy is finite in
$\lambda\rightarrow 0$ limit for both of the type 1 and 2 edges, which
nicely fits the observation of the edge
states for $k_\parallel=0$ for both types of edges in
Fig.~\ref{fig3}. On the other hand, for $k_\parallel=\pi$, the
entanglement entropy in $\lambda\rightarrow0$ limit is finite only for
the type 2 edge, and zero for the type 1 edge. Again, this result fits
appearance (absence) of the edge state for 
$k_\parallel=\pi$ for the type 1 (type 2) edge in
Fig.~\ref{fig3}. Figs.~\ref{fig4}(g) and
\ref{fig4}(h) also indicate that the entanglement entropy in
$\lambda\rightarrow 0$ limit is
really derived by the formula Eq.~\eqref{EE_formula} with extra factor
of two coming from our procedure to make bipartition in which two local
objects are broken.

To summarize, we develop a general theory to characterize a
gapped and short-range entangled state on the basis of adiabatic
continuation and the Berry phase. There, we give a natural
interpretation of the bulk--edge correspondence with the idea of a broken
local object. The relation between the Berry phase and the entanglement
entropy in a specific limit is also pointed out. In the latter
half, the general ideas are tested in a model with Dirac cones. We find
a new type of the Berry phase quantization, the quantization into
$\pm\pi/2$ in the introduced model. It is also shown that the new type
of the quantization modifies the way of edge state emergence from the
case of usual $0$/$\pi$ quantization.

\begin{acknowledgments}
 This work is partly supported by Grants-in-Aid for Scientific Research,
 No.26247064, No.23340112, No.25610101, and No.23540460 from JSPS.
\end{acknowledgments}

\bibliographystyle{apsrev4-1}
\bibliography{frac}

\begin{thebibliography}{38}%
\makeatletter
\providecommand \@ifxundefined [1]{%
 \@ifx{#1\undefined}
}%
\providecommand \@ifnum [1]{%
 \ifnum #1\expandafter \@firstoftwo
 \else \expandafter \@secondoftwo
 \fi
}%
\providecommand \@ifx [1]{%
 \ifx #1\expandafter \@firstoftwo
 \else \expandafter \@secondoftwo
 \fi
}%
\providecommand \natexlab [1]{#1}%
\providecommand \enquote  [1]{``#1''}%
\providecommand \bibnamefont  [1]{#1}%
\providecommand \bibfnamefont [1]{#1}%
\providecommand \citenamefont [1]{#1}%
\providecommand \href@noop [0]{\@secondoftwo}%
\providecommand \href [0]{\begingroup \@sanitize@url \@href}%
\providecommand \@href[1]{\@@startlink{#1}\@@href}%
\providecommand \@@href[1]{\endgroup#1\@@endlink}%
\providecommand \@sanitize@url [0]{\catcode `\\12\catcode `\$12\catcode
  `\&12\catcode `\#12\catcode `\^12\catcode `\_12\catcode `\%12\relax}%
\providecommand \@@startlink[1]{}%
\providecommand \@@endlink[0]{}%
\providecommand \url  [0]{\begingroup\@sanitize@url \@url }%
\providecommand \@url [1]{\endgroup\@href {#1}{\urlprefix }}%
\providecommand \urlprefix  [0]{URL }%
\providecommand \Eprint [0]{\href }%
\providecommand \doibase [0]{http://dx.doi.org/}%
\providecommand \selectlanguage [0]{\@gobble}%
\providecommand \bibinfo  [0]{\@secondoftwo}%
\providecommand \bibfield  [0]{\@secondoftwo}%
\providecommand \translation [1]{[#1]}%
\providecommand \BibitemOpen [0]{}%
\providecommand \bibitemStop [0]{}%
\providecommand \bibitemNoStop [0]{.\EOS\space}%
\providecommand \EOS [0]{\spacefactor3000\relax}%
\providecommand \BibitemShut  [1]{\csname bibitem#1\endcsname}%
\let\auto@bib@innerbib\@empty
%</preamble>
\bibitem [{\citenamefont {Wen}(1989)}]{PhysRevB.40.7387}%
  \BibitemOpen
  \bibfield  {author} {\bibinfo {author} {\bibfnamefont {X.~G.}\ \bibnamefont
  {Wen}},\ }\href {\doibase 10.1103/PhysRevB.40.7387} {\bibfield  {journal}
  {\bibinfo  {journal} {Phys. Rev. B}\ }\textbf {\bibinfo {volume} {40}},\
  \bibinfo {pages} {7387} (\bibinfo {year} {1989})}\BibitemShut {NoStop}%
\bibitem [{\citenamefont {Halperin}(1982)}]{PhysRevB.25.2185}%
  \BibitemOpen
  \bibfield  {author} {\bibinfo {author} {\bibfnamefont {B.~I.}\ \bibnamefont
  {Halperin}},\ }\href {\doibase 10.1103/PhysRevB.25.2185} {\bibfield
  {journal} {\bibinfo  {journal} {Phys. Rev. B}\ }\textbf {\bibinfo {volume}
  {25}},\ \bibinfo {pages} {2185} (\bibinfo {year} {1982})}\BibitemShut
  {NoStop}%
\bibitem [{\citenamefont {Su}\ \emph {et~al.}(1979)\citenamefont {Su},
  \citenamefont {Schrieffer},\ and\ \citenamefont
  {Heeger}}]{PhysRevLett.42.1698}%
  \BibitemOpen
  \bibfield  {author} {\bibinfo {author} {\bibfnamefont {W.~P.}\ \bibnamefont
  {Su}}, \bibinfo {author} {\bibfnamefont {J.~R.}\ \bibnamefont {Schrieffer}},
  \ and\ \bibinfo {author} {\bibfnamefont {A.~J.}\ \bibnamefont {Heeger}},\
  }\href {\doibase 10.1103/PhysRevLett.42.1698} {\bibfield  {journal} {\bibinfo
   {journal} {Phys. Rev. Lett.}\ }\textbf {\bibinfo {volume} {42}},\ \bibinfo
  {pages} {1698} (\bibinfo {year} {1979})}\BibitemShut {NoStop}%
\bibitem [{\citenamefont {Hu}(1994)}]{PhysRevLett.72.1526}%
  \BibitemOpen
  \bibfield  {author} {\bibinfo {author} {\bibfnamefont {C.-R.}\ \bibnamefont
  {Hu}},\ }\href {\doibase 10.1103/PhysRevLett.72.1526} {\bibfield  {journal}
  {\bibinfo  {journal} {Phys. Rev. Lett.}\ }\textbf {\bibinfo {volume} {72}},\
  \bibinfo {pages} {1526} (\bibinfo {year} {1994})}\BibitemShut {NoStop}%
\bibitem [{\citenamefont {Tanaka}\ and\ \citenamefont
  {Kashiwaya}(1995)}]{PhysRevLett.74.3451}%
  \BibitemOpen
  \bibfield  {author} {\bibinfo {author} {\bibfnamefont {Y.}~\bibnamefont
  {Tanaka}}\ and\ \bibinfo {author} {\bibfnamefont {S.}~\bibnamefont
  {Kashiwaya}},\ }\href {\doibase 10.1103/PhysRevLett.74.3451} {\bibfield
  {journal} {\bibinfo  {journal} {Phys. Rev. Lett.}\ }\textbf {\bibinfo
  {volume} {74}},\ \bibinfo {pages} {3451} (\bibinfo {year}
  {1995})}\BibitemShut {NoStop}%
\bibitem [{\citenamefont {Fujita}\ \emph {et~al.}(1996)\citenamefont {Fujita},
  \citenamefont {Wakabayashi}, \citenamefont {Nakada},\ and\ \citenamefont
  {Kusakabe}}]{JPSJ.65.1920}%
  \BibitemOpen
  \bibfield  {author} {\bibinfo {author} {\bibfnamefont {M.}~\bibnamefont
  {Fujita}}, \bibinfo {author} {\bibfnamefont {K.}~\bibnamefont {Wakabayashi}},
  \bibinfo {author} {\bibfnamefont {K.}~\bibnamefont {Nakada}}, \ and\ \bibinfo
  {author} {\bibfnamefont {K.}~\bibnamefont {Kusakabe}},\ }\href@noop {}
  {\bibfield  {journal} {\bibinfo  {journal} {J. Phys. Soc. Jpn.}\ }\textbf
  {\bibinfo {volume} {65}},\ \bibinfo {pages} {1920} (\bibinfo {year}
  {1996})}\BibitemShut {NoStop}%
\bibitem [{\citenamefont {Kitaev}(2001)}]{1063-7869-44-10S-S29}%
  \BibitemOpen
  \bibfield  {author} {\bibinfo {author} {\bibfnamefont {A.~Y.}\ \bibnamefont
  {Kitaev}},\ }\href {http://stacks.iop.org/1063-7869/44/i=10S/a=S29}
  {\bibfield  {journal} {\bibinfo  {journal} {Physics-Uspekhi}\ }\textbf
  {\bibinfo {volume} {44}},\ \bibinfo {pages} {131} (\bibinfo {year}
  {2001})}\BibitemShut {NoStop}%
\bibitem [{\citenamefont {Volovik}(2003)}]{volovik}%
  \BibitemOpen
  \bibfield  {author} {\bibinfo {author} {\bibnamefont {Volovik}},\ }\href@noop
  {} {\emph {\bibinfo {title} {The Universe in a Helium Droplet}}}\ (\bibinfo
  {publisher} {Clarendon Press, Oxford},\ \bibinfo {year} {2003})\BibitemShut
  {NoStop}%
\bibitem [{\citenamefont {Thouless}\ \emph {et~al.}(1982)\citenamefont
  {Thouless}, \citenamefont {Kohmoto}, \citenamefont {Nightingale},\ and\
  \citenamefont {den Nijs}}]{PhysRevLett.49.405}%
  \BibitemOpen
  \bibfield  {author} {\bibinfo {author} {\bibfnamefont {D.~J.}\ \bibnamefont
  {Thouless}}, \bibinfo {author} {\bibfnamefont {M.}~\bibnamefont {Kohmoto}},
  \bibinfo {author} {\bibfnamefont {M.~P.}\ \bibnamefont {Nightingale}}, \ and\
  \bibinfo {author} {\bibfnamefont {M.}~\bibnamefont {den Nijs}},\ }\href
  {\doibase 10.1103/PhysRevLett.49.405} {\bibfield  {journal} {\bibinfo
  {journal} {Phys. Rev. Lett.}\ }\textbf {\bibinfo {volume} {49}},\ \bibinfo
  {pages} {405} (\bibinfo {year} {1982})}\BibitemShut {NoStop}%
\bibitem [{\citenamefont {Hatsugai}(1993{\natexlab{a}})}]{PhysRevLett.71.3697}%
  \BibitemOpen
  \bibfield  {author} {\bibinfo {author} {\bibfnamefont {Y.}~\bibnamefont
  {Hatsugai}},\ }\href {\doibase 10.1103/PhysRevLett.71.3697} {\bibfield
  {journal} {\bibinfo  {journal} {Phys. Rev. Lett.}\ }\textbf {\bibinfo
  {volume} {71}},\ \bibinfo {pages} {3697} (\bibinfo {year}
  {1993}{\natexlab{a}})}\BibitemShut {NoStop}%
\bibitem [{\citenamefont {Hatsugai}(1993{\natexlab{b}})}]{PhysRevB.48.11851}%
  \BibitemOpen
  \bibfield  {author} {\bibinfo {author} {\bibfnamefont {Y.}~\bibnamefont
  {Hatsugai}},\ }\href {\doibase 10.1103/PhysRevB.48.11851} {\bibfield
  {journal} {\bibinfo  {journal} {Phys. Rev. B}\ }\textbf {\bibinfo {volume}
  {48}},\ \bibinfo {pages} {11851} (\bibinfo {year}
  {1993}{\natexlab{b}})}\BibitemShut {NoStop}%
\bibitem [{\citenamefont {Hasan}\ and\ \citenamefont
  {Kane}(2010)}]{RevModPhys.82.3045}%
  \BibitemOpen
  \bibfield  {author} {\bibinfo {author} {\bibfnamefont {M.~Z.}\ \bibnamefont
  {Hasan}}\ and\ \bibinfo {author} {\bibfnamefont {C.~L.}\ \bibnamefont
  {Kane}},\ }\href {\doibase 10.1103/RevModPhys.82.3045} {\bibfield  {journal}
  {\bibinfo  {journal} {Rev. Mod. Phys.}\ }\textbf {\bibinfo {volume} {82}},\
  \bibinfo {pages} {3045} (\bibinfo {year} {2010})}\BibitemShut {NoStop}%
\bibitem [{\citenamefont {Qi}\ and\ \citenamefont
  {Zhang}(2011)}]{RevModPhys.83.1057}%
  \BibitemOpen
  \bibfield  {author} {\bibinfo {author} {\bibfnamefont {X.-L.}\ \bibnamefont
  {Qi}}\ and\ \bibinfo {author} {\bibfnamefont {S.-C.}\ \bibnamefont {Zhang}},\
  }\href {\doibase 10.1103/RevModPhys.83.1057} {\bibfield  {journal} {\bibinfo
  {journal} {Rev. Mod. Phys.}\ }\textbf {\bibinfo {volume} {83}},\ \bibinfo
  {pages} {1057} (\bibinfo {year} {2011})}\BibitemShut {NoStop}%
\bibitem [{\citenamefont {Ryu}\ and\ \citenamefont
  {Hatsugai}(2002)}]{PhysRevLett.89.077002}%
  \BibitemOpen
  \bibfield  {author} {\bibinfo {author} {\bibfnamefont {S.}~\bibnamefont
  {Ryu}}\ and\ \bibinfo {author} {\bibfnamefont {Y.}~\bibnamefont {Hatsugai}},\
  }\href@noop {} {\bibfield  {journal} {\bibinfo  {journal} {Phys. Rev. Lett.}\
  }\textbf {\bibinfo {volume} {89}},\ \bibinfo {pages} {077002} (\bibinfo
  {year} {2002})}\BibitemShut {NoStop}%
\bibitem [{\citenamefont {Berry}(1984)}]{Berry84}%
  \BibitemOpen
  \bibfield  {author} {\bibinfo {author} {\bibfnamefont {M.~V.}\ \bibnamefont
  {Berry}},\ }\href@noop {} {\bibfield  {journal} {\bibinfo  {journal} {Proc.\
  R.\ Soc.}\ }\textbf {\bibinfo {volume} {A392}},\ \bibinfo {pages} {45}
  (\bibinfo {year} {1984})}\BibitemShut {NoStop}%
\bibitem [{\citenamefont {Hatsugai}(2004)}]{doi:10.1143/JPSJ.73.2604}%
  \BibitemOpen
  \bibfield  {author} {\bibinfo {author} {\bibfnamefont {Y.}~\bibnamefont
  {Hatsugai}},\ }\href {\doibase 10.1143/JPSJ.73.2604} {\bibfield  {journal}
  {\bibinfo  {journal} {J. Phys. Soc. Jpn.}\ }\textbf {\bibinfo {volume}
  {73}},\ \bibinfo {pages} {2604} (\bibinfo {year} {2004})}\BibitemShut
  {NoStop}%
\bibitem [{\citenamefont {Hatsugai}(2005)}]{doi:10.1143/JPSJ.74.1374}%
  \BibitemOpen
  \bibfield  {author} {\bibinfo {author} {\bibfnamefont {Y.}~\bibnamefont
  {Hatsugai}},\ }\href {\doibase 10.1143/JPSJ.74.1374} {\bibfield  {journal}
  {\bibinfo  {journal} {J. Phys. Soc. Jpn.}\ }\textbf {\bibinfo {volume}
  {74}},\ \bibinfo {pages} {1374} (\bibinfo {year} {2005})}\BibitemShut
  {NoStop}%
\bibitem [{\citenamefont {Hatsugai}(2006)}]{doi:10.1143/JPSJ.75.123601}%
  \BibitemOpen
  \bibfield  {author} {\bibinfo {author} {\bibfnamefont {Y.}~\bibnamefont
  {Hatsugai}},\ }\href {\doibase 10.1143/JPSJ.75.123601} {\bibfield  {journal}
  {\bibinfo  {journal} {J. Phys. Soc. Jpn.}\ }\textbf {\bibinfo {volume}
  {75}},\ \bibinfo {pages} {123601} (\bibinfo {year} {2006})}\BibitemShut
  {NoStop}%
\bibitem [{\citenamefont {Hatsugai}(2010)}]{Hatsugai10Z2}%
  \BibitemOpen
  \bibfield  {author} {\bibinfo {author} {\bibfnamefont {Y.}~\bibnamefont
  {Hatsugai}},\ }\href {http://stacks.iop.org/1367-2630/12/i=6/a=065004}
  {\bibfield  {journal} {\bibinfo  {journal} {New J. Phys.}\ }\textbf {\bibinfo
  {volume} {12}},\ \bibinfo {pages} {065004} (\bibinfo {year}
  {2010})}\BibitemShut {NoStop}%
\bibitem [{\citenamefont {Kariyado}\ and\ \citenamefont
  {Hatsugai}(2013)}]{PhysRevB.88.245126}%
  \BibitemOpen
  \bibfield  {author} {\bibinfo {author} {\bibfnamefont {T.}~\bibnamefont
  {Kariyado}}\ and\ \bibinfo {author} {\bibfnamefont {Y.}~\bibnamefont
  {Hatsugai}},\ }\href@noop {} {\bibfield  {journal} {\bibinfo  {journal}
  {Phys. Rev. B}\ }\textbf {\bibinfo {volume} {88}},\ \bibinfo {pages} {245126}
  (\bibinfo {year} {2013})}\BibitemShut {NoStop}%
\bibitem [{\citenamefont {Ryu}\ and\ \citenamefont
  {Hatsugai}(2006)}]{PhysRevB.73.245115}%
  \BibitemOpen
  \bibfield  {author} {\bibinfo {author} {\bibfnamefont {S.}~\bibnamefont
  {Ryu}}\ and\ \bibinfo {author} {\bibfnamefont {Y.}~\bibnamefont {Hatsugai}},\
  }\href {\doibase 10.1103/PhysRevB.73.245115} {\bibfield  {journal} {\bibinfo
  {journal} {Phys. Rev. B}\ }\textbf {\bibinfo {volume} {73}},\ \bibinfo
  {pages} {245115} (\bibinfo {year} {2006})}\BibitemShut {NoStop}%
\bibitem [{\citenamefont {Arikawa}\ \emph {et~al.}(2009)\citenamefont
  {Arikawa}, \citenamefont {Tanaya}, \citenamefont {Maruyama},\ and\
  \citenamefont {Hatsugai}}]{PhysRevB.79.205107}%
  \BibitemOpen
  \bibfield  {author} {\bibinfo {author} {\bibfnamefont {M.}~\bibnamefont
  {Arikawa}}, \bibinfo {author} {\bibfnamefont {S.}~\bibnamefont {Tanaya}},
  \bibinfo {author} {\bibfnamefont {I.}~\bibnamefont {Maruyama}}, \ and\
  \bibinfo {author} {\bibfnamefont {Y.}~\bibnamefont {Hatsugai}},\ }\href
  {\doibase 10.1103/PhysRevB.79.205107} {\bibfield  {journal} {\bibinfo
  {journal} {Phys. Rev. B}\ }\textbf {\bibinfo {volume} {79}},\ \bibinfo
  {pages} {205107} (\bibinfo {year} {2009})}\BibitemShut {NoStop}%
\bibitem [{\citenamefont {Chung}\ \emph {et~al.}(2011)\citenamefont {Chung},
  \citenamefont {Jhu}, \citenamefont {Chen},\ and\ \citenamefont
  {Yip}}]{0295-5075-95-2-27003}%
  \BibitemOpen
  \bibfield  {author} {\bibinfo {author} {\bibfnamefont {M.-C.}\ \bibnamefont
  {Chung}}, \bibinfo {author} {\bibfnamefont {Y.-H.}\ \bibnamefont {Jhu}},
  \bibinfo {author} {\bibfnamefont {P.}~\bibnamefont {Chen}}, \ and\ \bibinfo
  {author} {\bibfnamefont {S.}~\bibnamefont {Yip}},\ }\href
  {http://stacks.iop.org/0295-5075/95/i=2/a=27003} {\bibfield  {journal}
  {\bibinfo  {journal} {Europhys. Lett.}\ }\textbf {\bibinfo {volume} {95}},\
  \bibinfo {pages} {27003} (\bibinfo {year} {2011})}\BibitemShut {NoStop}%
\bibitem [{\citenamefont {Hughes}\ \emph {et~al.}(2011)\citenamefont {Hughes},
  \citenamefont {Prodan},\ and\ \citenamefont {Bernevig}}]{PhysRevB.83.245132}%
  \BibitemOpen
  \bibfield  {author} {\bibinfo {author} {\bibfnamefont {T.~L.}\ \bibnamefont
  {Hughes}}, \bibinfo {author} {\bibfnamefont {E.}~\bibnamefont {Prodan}}, \
  and\ \bibinfo {author} {\bibfnamefont {B.~A.}\ \bibnamefont {Bernevig}},\
  }\href {\doibase 10.1103/PhysRevB.83.245132} {\bibfield  {journal} {\bibinfo
  {journal} {Phys. Rev. B}\ }\textbf {\bibinfo {volume} {83}},\ \bibinfo
  {pages} {245132} (\bibinfo {year} {2011})}\BibitemShut {NoStop}%
\bibitem [{\citenamefont {Chandran}\ \emph {et~al.}(2011)\citenamefont
  {Chandran}, \citenamefont {Hermanns}, \citenamefont {Regnault},\ and\
  \citenamefont {Bernevig}}]{PhysRevB.84.205136}%
  \BibitemOpen
  \bibfield  {author} {\bibinfo {author} {\bibfnamefont {A.}~\bibnamefont
  {Chandran}}, \bibinfo {author} {\bibfnamefont {M.}~\bibnamefont {Hermanns}},
  \bibinfo {author} {\bibfnamefont {N.}~\bibnamefont {Regnault}}, \ and\
  \bibinfo {author} {\bibfnamefont {B.~A.}\ \bibnamefont {Bernevig}},\ }\href
  {\doibase 10.1103/PhysRevB.84.205136} {\bibfield  {journal} {\bibinfo
  {journal} {Phys. Rev. B}\ }\textbf {\bibinfo {volume} {84}},\ \bibinfo
  {pages} {205136} (\bibinfo {year} {2011})}\BibitemShut {NoStop}%
\bibitem [{\citenamefont {Chen}\ \emph {et~al.}(2010)\citenamefont {Chen},
  \citenamefont {Gu},\ and\ \citenamefont {Wen}}]{PhysRevB.82.155138}%
  \BibitemOpen
  \bibfield  {author} {\bibinfo {author} {\bibfnamefont {X.}~\bibnamefont
  {Chen}}, \bibinfo {author} {\bibfnamefont {Z.-C.}\ \bibnamefont {Gu}}, \ and\
  \bibinfo {author} {\bibfnamefont {X.-G.}\ \bibnamefont {Wen}},\ }\href
  {\doibase 10.1103/PhysRevB.82.155138} {\bibfield  {journal} {\bibinfo
  {journal} {Phys. Rev. B}\ }\textbf {\bibinfo {volume} {82}},\ \bibinfo
  {pages} {155138} (\bibinfo {year} {2010})}\BibitemShut {NoStop}%
\bibitem [{\citenamefont {Pollmann}\ \emph {et~al.}(2012)\citenamefont
  {Pollmann}, \citenamefont {Berg}, \citenamefont {Turner},\ and\ \citenamefont
  {Oshikawa}}]{PhysRevB.85.075125}%
  \BibitemOpen
  \bibfield  {author} {\bibinfo {author} {\bibfnamefont {F.}~\bibnamefont
  {Pollmann}}, \bibinfo {author} {\bibfnamefont {E.}~\bibnamefont {Berg}},
  \bibinfo {author} {\bibfnamefont {A.~M.}\ \bibnamefont {Turner}}, \ and\
  \bibinfo {author} {\bibfnamefont {M.}~\bibnamefont {Oshikawa}},\ }\href
  {\doibase 10.1103/PhysRevB.85.075125} {\bibfield  {journal} {\bibinfo
  {journal} {Phys. Rev. B}\ }\textbf {\bibinfo {volume} {85}},\ \bibinfo
  {pages} {075125} (\bibinfo {year} {2012})}\BibitemShut {NoStop}%
\bibitem [{\citenamefont {Hirano}\ \emph
  {et~al.}(2008{\natexlab{a}})\citenamefont {Hirano}, \citenamefont {Katsura},\
  and\ \citenamefont {Hatsugai}}]{PhysRevB.78.054431}%
  \BibitemOpen
  \bibfield  {author} {\bibinfo {author} {\bibfnamefont {T.}~\bibnamefont
  {Hirano}}, \bibinfo {author} {\bibfnamefont {H.}~\bibnamefont {Katsura}}, \
  and\ \bibinfo {author} {\bibfnamefont {Y.}~\bibnamefont {Hatsugai}},\ }\href
  {\doibase 10.1103/PhysRevB.78.054431} {\bibfield  {journal} {\bibinfo
  {journal} {Phys. Rev. B}\ }\textbf {\bibinfo {volume} {78}},\ \bibinfo
  {pages} {054431} (\bibinfo {year} {2008}{\natexlab{a}})}\BibitemShut
  {NoStop}%
\bibitem [{\citenamefont {Maruyama}\ and\ \citenamefont
  {Hatsugai}(2007)}]{doi:10.1143/JPSJ.76.113601}%
  \BibitemOpen
  \bibfield  {author} {\bibinfo {author} {\bibfnamefont {I.}~\bibnamefont
  {Maruyama}}\ and\ \bibinfo {author} {\bibfnamefont {Y.}~\bibnamefont
  {Hatsugai}},\ }\href {\doibase 10.1143/JPSJ.76.113601} {\bibfield  {journal}
  {\bibinfo  {journal} {J. Phys. Soc. Jpn.}\ }\textbf {\bibinfo {volume}
  {76}},\ \bibinfo {pages} {113601} (\bibinfo {year} {2007})}\BibitemShut
  {NoStop}%
\bibitem [{\citenamefont {Hirano}\ \emph
  {et~al.}(2008{\natexlab{b}})\citenamefont {Hirano}, \citenamefont {Katsura},\
  and\ \citenamefont {Hatsugai}}]{PhysRevB.77.094431}%
  \BibitemOpen
  \bibfield  {author} {\bibinfo {author} {\bibfnamefont {T.}~\bibnamefont
  {Hirano}}, \bibinfo {author} {\bibfnamefont {H.}~\bibnamefont {Katsura}}, \
  and\ \bibinfo {author} {\bibfnamefont {Y.}~\bibnamefont {Hatsugai}},\ }\href
  {\doibase 10.1103/PhysRevB.77.094431} {\bibfield  {journal} {\bibinfo
  {journal} {Phys. Rev. B}\ }\textbf {\bibinfo {volume} {77}},\ \bibinfo
  {pages} {094431} (\bibinfo {year} {2008}{\natexlab{b}})}\BibitemShut
  {NoStop}%
\bibitem [{\citenamefont {Maruyama}\ \emph {et~al.}(2009)\citenamefont
  {Maruyama}, \citenamefont {Hirano},\ and\ \citenamefont
  {Hatsugai}}]{PhysRevB.79.115107}%
  \BibitemOpen
  \bibfield  {author} {\bibinfo {author} {\bibfnamefont {I.}~\bibnamefont
  {Maruyama}}, \bibinfo {author} {\bibfnamefont {T.}~\bibnamefont {Hirano}}, \
  and\ \bibinfo {author} {\bibfnamefont {Y.}~\bibnamefont {Hatsugai}},\ }\href
  {\doibase 10.1103/PhysRevB.79.115107} {\bibfield  {journal} {\bibinfo
  {journal} {Phys. Rev. B}\ }\textbf {\bibinfo {volume} {79}},\ \bibinfo
  {pages} {115107} (\bibinfo {year} {2009})}\BibitemShut {NoStop}%
\bibitem [{foo()}]{footnote1}%
  \BibitemOpen
  \href@noop {} {}\bibinfo {howpublished} {When the boundary allows multiple
  edge states within a finite region, residual interaction among the edge
  states can be effective which makes the edge states gapped.}\BibitemShut
  {Stop}%
\bibitem [{\citenamefont {Hatsugai}(2009)}]{Hatsugai20091061}%
  \BibitemOpen
  \bibfield  {author} {\bibinfo {author} {\bibfnamefont {Y.}~\bibnamefont
  {Hatsugai}},\ }\href@noop {} {\bibfield  {journal} {\bibinfo  {journal}
  {Solid State Commun.}\ }\textbf {\bibinfo {volume} {149}},\ \bibinfo {pages}
  {1061 } (\bibinfo {year} {2009})}\BibitemShut {NoStop}%
\bibitem [{\citenamefont {Delplace}\ \emph {et~al.}(2011)\citenamefont
  {Delplace}, \citenamefont {Ullmo},\ and\ \citenamefont
  {Montambaux}}]{PhysRevB.84.195452}%
  \BibitemOpen
  \bibfield  {author} {\bibinfo {author} {\bibfnamefont {P.}~\bibnamefont
  {Delplace}}, \bibinfo {author} {\bibfnamefont {D.}~\bibnamefont {Ullmo}}, \
  and\ \bibinfo {author} {\bibfnamefont {G.}~\bibnamefont {Montambaux}},\
  }\href {\doibase 10.1103/PhysRevB.84.195452} {\bibfield  {journal} {\bibinfo
  {journal} {Phys. Rev. B}\ }\textbf {\bibinfo {volume} {84}},\ \bibinfo
  {pages} {195452} (\bibinfo {year} {2011})}\BibitemShut {NoStop}%
\bibitem [{\citenamefont {Yamashita}\ \emph {et~al.}(2013)\citenamefont
  {Yamashita}, \citenamefont {Tomura}, \citenamefont {Yanagi},\ and\
  \citenamefont {Ueda}}]{PhysRevB.88.195104}%
  \BibitemOpen
  \bibfield  {author} {\bibinfo {author} {\bibfnamefont {Y.}~\bibnamefont
  {Yamashita}}, \bibinfo {author} {\bibfnamefont {M.}~\bibnamefont {Tomura}},
  \bibinfo {author} {\bibfnamefont {Y.}~\bibnamefont {Yanagi}}, \ and\ \bibinfo
  {author} {\bibfnamefont {K.}~\bibnamefont {Ueda}},\ }\href {\doibase
  10.1103/PhysRevB.88.195104} {\bibfield  {journal} {\bibinfo  {journal} {Phys.
  Rev. B}\ }\textbf {\bibinfo {volume} {88}},\ \bibinfo {pages} {195104}
  (\bibinfo {year} {2013})}\BibitemShut {NoStop}%
\bibitem [{\citenamefont {Zak}(1989)}]{PhysRevLett.62.2747}%
  \BibitemOpen
  \bibfield  {author} {\bibinfo {author} {\bibfnamefont {J.}~\bibnamefont
  {Zak}},\ }\href {\doibase 10.1103/PhysRevLett.62.2747} {\bibfield  {journal}
  {\bibinfo  {journal} {Phys. Rev. Lett.}\ }\textbf {\bibinfo {volume} {62}},\
  \bibinfo {pages} {2747} (\bibinfo {year} {1989})}\BibitemShut {NoStop}%
\bibitem [{\citenamefont {King-Smith}\ and\ \citenamefont
  {Vanderbilt}(1993)}]{PhysRevB.47.1651}%
  \BibitemOpen
  \bibfield  {author} {\bibinfo {author} {\bibfnamefont {R.~D.}\ \bibnamefont
  {King-Smith}}\ and\ \bibinfo {author} {\bibfnamefont {D.}~\bibnamefont
  {Vanderbilt}},\ }\href {\doibase 10.1103/PhysRevB.47.1651} {\bibfield
  {journal} {\bibinfo  {journal} {Phys. Rev. B}\ }\textbf {\bibinfo {volume}
  {47}},\ \bibinfo {pages} {1651} (\bibinfo {year} {1993})}\BibitemShut
  {NoStop}%
\bibitem [{\citenamefont {Peschel}\ and\ \citenamefont
  {Eisler}(2009)}]{1751-8121-42-50-504003}%
  \BibitemOpen
  \bibfield  {author} {\bibinfo {author} {\bibfnamefont {I.}~\bibnamefont
  {Peschel}}\ and\ \bibinfo {author} {\bibfnamefont {V.}~\bibnamefont
  {Eisler}},\ }\href {http://stacks.iop.org/1751-8121/42/i=50/a=504003}
  {\bibfield  {journal} {\bibinfo  {journal} {J. Phys. A: Math. Theor.}\
  }\textbf {\bibinfo {volume} {42}},\ \bibinfo {pages} {504003} (\bibinfo
  {year} {2009})}\BibitemShut {NoStop}%
\end{thebibliography}

\end{document}